\documentclass[preprint]{aastex631}
\reportnum{DES-2022-0737}
\reportnum{FERMILAB-FN-1227}

\begin{document}
\title{Cool Cores in Clusters of Galaxies in the Dark Energy Survey}

\author{K.~Graham}
\affiliation{University of California, Santa Cruz, Santa Cruz, CA 95064, USA}
\affiliation{Santa Cruz Institute for Particle Physics, Santa Cruz, CA 95064, USA}
\author{J.~O'Donnell}
\affiliation{University of California, Santa Cruz, Santa Cruz, CA 95064, USA}
\affiliation{Santa Cruz Institute for Particle Physics, Santa Cruz, CA 95064, USA}
\author{M.~M.~Silverstein}
\affiliation{University of California, Santa Cruz, Santa Cruz, CA 95064, USA}
\affiliation{Santa Cruz Institute for Particle Physics, Santa Cruz, CA 95064, USA}
\author{O.~Eiger}
\affiliation{SLAC National Accelerator Laboratory, 2575 Sand Hill Road, Menlo Park, CA 94025}
\affiliation{University of California, Santa Cruz, Santa Cruz, CA 95064, USA}
\affiliation{Santa Cruz Institute for Particle Physics, Santa Cruz, CA 95064, USA}
\author{T.~E.~Jeltema}
\affiliation{University of California, Santa Cruz, Santa Cruz, CA 95064, USA}
\affiliation{Santa Cruz Institute for Particle Physics, Santa Cruz, CA 95064, USA}
\author{D.~L.~Hollowood}
\affiliation{University of California, Santa Cruz, Santa Cruz, CA 95064, USA}
\affiliation{Santa Cruz Institute for Particle Physics, Santa Cruz, CA 95064, USA}
\author{D.~Cross}
\affiliation{Institute of Space Sciences (ICE, CSIC),  Campus UAB, Carrer de Can Magrans, s/n,  08193 Barcelona, Spain}
\author{S.~Everett}
\affiliation{Jet Propulsion Laboratory, California Institute of Technology, 4800 Oak Grove Dr., Pasadena, CA 91109, USA}
\author{P.~Giles}
\affiliation{Department of Physics and Astronomy, Pevensey Building, University of Sussex, Brighton, BN1 9QH, UK}
\author{J.~Jobel}
\affiliation{University of California, Santa Cruz, Santa Cruz, CA 95064, USA}
\affiliation{Santa Cruz Institute for Particle Physics, Santa Cruz, CA 95064, USA}
\author{D.~Laubner}
\affiliation{University of California, Santa Cruz, Santa Cruz, CA 95064, USA}
\affiliation{Santa Cruz Institute for Particle Physics, Santa Cruz, CA 95064, USA}
\author{A.~McDaniel}
\affiliation{Department of Physics and Astronomy, Clemson University, Kinard Lab of Physics, Clemson, SC 29634-0978, US}
\author{A.~K.~Romer}
\affiliation{Department of Physics and Astronomy, Pevensey Building, University of Sussex, Brighton, BN1 9QH, UK}
\author{A.~Swart}
\affiliation{San Francisco State University, 1600 Holloway Ave, San Francisco, CA 94132}
\author{M.~Aguena}
\affiliation{Laborat\'orio Interinstitucional de e-Astronomia - LIneA, Rua Gal. Jos\'e Cristino 77, Rio de Janeiro, RJ - 20921-400, Brazil}
\author{S.~Allam}
\affiliation{Fermi National Accelerator Laboratory, P. O. Box 500, Batavia, IL 60510, USA}
\author{O.~Alves}
\affiliation{Department of Physics, University of Michigan, Ann Arbor, MI 48109, USA}
\author{D.~Brooks}
\affiliation{Department of Physics \& Astronomy, University College London, Gower Street, London, WC1E 6BT, UK}
\author{M.~Carrasco~Kind}
\affiliation{Center for Astrophysical Surveys, National Center for Supercomputing Applications, 1205 West Clark St., Urbana, IL 61801, USA}
\affiliation{Department of Astronomy, University of Illinois at Urbana-Champaign, 1002 W. Green Street, Urbana, IL 61801, USA}
\author{J.~Carretero}
\affiliation{Institut de F\'{\i}sica d'Altes Energies (IFAE), The Barcelona Institute of Science and Technology, Campus UAB, 08193 Bellaterra (Barcelona) Spain}
\author{M.~Costanzi}
\affiliation{Astronomy Unit, Department of Physics, University of Trieste, via Tiepolo 11, I-34131 Trieste, Italy}
\affiliation{INAF-Osservatorio Astronomico di Trieste, via G. B. Tiepolo 11, I-34143 Trieste, Italy}
\affiliation{Institute for Fundamental Physics of the Universe, Via Beirut 2, 34014 Trieste, Italy}
\author{L.~N.~da Costa}
\affiliation{Laborat\'orio Interinstitucional de e-Astronomia - LIneA, Rua Gal. Jos\'e Cristino 77, Rio de Janeiro, RJ - 20921-400, Brazil}
\author{M.~E.~S.~Pereira}
\affiliation{Hamburger Sternwarte, Universit\"{a}t Hamburg, Gojenbergsweg 112, 21029 Hamburg, Germany}
\author{J.~De~Vicente}
\affiliation{Centro de Investigaciones Energ\'eticas, Medioambientales y Tecnol\'ogicas (CIEMAT), Madrid, Spain}
\author{S.~Desai}
\affiliation{Department of Physics, IIT Hyderabad, Kandi, Telangana 502285, India}
\author{J.~P.~Dietrich}
\affiliation{University Observatory, Faculty of Physics, Ludwig-Maximilians-Universit\"at, Scheinerstr. 1, 81679 Munich, Germany}
\author{P.~Doel}
\affiliation{Department of Physics \& Astronomy, University College London, Gower Street, London, WC1E 6BT, UK}
\author{I.~Ferrero}
\affiliation{Institute of Theoretical Astrophysics, University of Oslo. P.O. Box 1029 Blindern, NO-0315 Oslo, Norway}
\author{J.~Frieman}
\affiliation{Fermi National Accelerator Laboratory, P. O. Box 500, Batavia, IL 60510, USA}
\affiliation{Kavli Institute for Cosmological Physics, University of Chicago, Chicago, IL 60637, USA}
\author{J.~Garc\'ia-Bellido}
\affiliation{Instituto de Fisica Teorica UAM/CSIC, Universidad Autonoma de Madrid, 28049 Madrid, Spain}
\author{D.~Gruen}
\affiliation{University Observatory, Faculty of Physics, Ludwig-Maximilians-Universit\"at, Scheinerstr. 1, 81679 Munich, Germany}
\author{R.~A.~Gruendl}
\affiliation{Center for Astrophysical Surveys, National Center for Supercomputing Applications, 1205 West Clark St., Urbana, IL 61801, USA}
\affiliation{Department of Astronomy, University of Illinois at Urbana-Champaign, 1002 W. Green Street, Urbana, IL 61801, USA}
\author{S.~R.~Hinton}
\affiliation{School of Mathematics and Physics, University of Queensland,  Brisbane, QLD 4072, Australia}
\author{K.~Honscheid}
\affiliation{Center for Cosmology and Astro-Particle Physics, The Ohio State University, Columbus, OH 43210, USA}
\affiliation{Department of Physics, The Ohio State University, Columbus, OH 43210, USA}
\author{D.~J.~James}
\affiliation{Center for Astrophysics $\vert$ Harvard \& Smithsonian, 60 Garden Street, Cambridge, MA 02138, USA}
\author{K.~Kuehn}
\affiliation{Australian Astronomical Optics, Macquarie University, North Ryde, NSW 2113, Australia}
\affiliation{Lowell Observatory, 1400 Mars Hill Rd, Flagstaff, AZ 86001, USA}
\author{N.~Kuropatkin}
\affiliation{Fermi National Accelerator Laboratory, P. O. Box 500, Batavia, IL 60510, USA}
\author{O.~Lahav}
\affiliation{Department of Physics \& Astronomy, University College London, Gower Street, London, WC1E 6BT, UK}
\author{J.~L.~Marshall}
\affiliation{George P. and Cynthia Woods Mitchell Institute for Fundamental Physics and Astronomy, and Department of Physics and Astronomy, Texas A\&M University, College Station, TX 77843,  USA}
\author{P.~Melchior}
\affiliation{Department of Astrophysical Sciences, Princeton University, Peyton Hall, Princeton, NJ 08544, USA}
\author{J. Mena-Fern{\'a}ndez}
\affiliation{Centro de Investigaciones Energ\'eticas, Medioambientales y Tecnol\'ogicas (CIEMAT), Madrid, Spain}
\author{F.~Menanteau}
\affiliation{Center for Astrophysical Surveys, National Center for Supercomputing Applications, 1205 West Clark St., Urbana, IL 61801, USA}
\affiliation{Department of Astronomy, University of Illinois at Urbana-Champaign, 1002 W. Green Street, Urbana, IL 61801, USA}
\author{R.~Miquel}
\affiliation{Instituci\'o Catalana de Recerca i Estudis Avan\c{c}ats, E-08010 Barcelona, Spain}
\affiliation{Institut de F\'{\i}sica d'Altes Energies (IFAE), The Barcelona Institute of Science and Technology, Campus UAB, 08193 Bellaterra (Barcelona) Spain}
\author{R.~L.~C.~Ogando}
\affiliation{Observat\'orio Nacional, Rua Gal. Jos\'e Cristino 77, Rio de Janeiro, RJ - 20921-400, Brazil}
\author{A.~Palmese}
\affiliation{Department of Physics, Carnegie Mellon University, Pittsburgh, Pennsylvania 15312, USA}
\author{A.~Pieres}
\affiliation{Laborat\'orio Interinstitucional de e-Astronomia - LIneA, Rua Gal. Jos\'e Cristino 77, Rio de Janeiro, RJ - 20921-400, Brazil}
\affiliation{Observat\'orio Nacional, Rua Gal. Jos\'e Cristino 77, Rio de Janeiro, RJ - 20921-400, Brazil}
\author{A.~A.~Plazas~Malag\'on}
\affiliation{Department of Astrophysical Sciences, Princeton University, Peyton Hall, Princeton, NJ 08544, USA}
\author{K.~Reil}
\affiliation{SLAC National Accelerator Laboratory, Menlo Park, CA 94025, USA}
\author{M.~Rodriguez-Monroy}
\affiliation{Centro de Investigaciones Energ\'eticas, Medioambientales y Tecnol\'ogicas (CIEMAT), Madrid, Spain}
\author{E.~Sanchez}
\affiliation{Centro de Investigaciones Energ\'eticas, Medioambientales y Tecnol\'ogicas (CIEMAT), Madrid, Spain}
\author{V.~Scarpine}
\affiliation{Fermi National Accelerator Laboratory, P. O. Box 500, Batavia, IL 60510, USA}
\author{M.~Schubnell}
\affiliation{Department of Physics, University of Michigan, Ann Arbor, MI 48109, USA}
\author{M.~Smith}
\affiliation{School of Physics and Astronomy, University of Southampton,  Southampton, SO17 1BJ, UK}
\author{E.~Suchyta}
\affiliation{Computer Science and Mathematics Division, Oak Ridge National Laboratory, Oak Ridge, TN 37831}
\author{G.~Tarle}
\affiliation{Department of Physics, University of Michigan, Ann Arbor, MI 48109, USA}
\author{C.~To}
\affiliation{Center for Cosmology and Astro-Particle Physics, The Ohio State University, Columbus, OH 43210, USA}
\author{N.~Weaverdyck}
\affiliation{Department of Physics, University of Michigan, Ann Arbor, MI 48109, USA}
\affiliation{Lawrence Berkeley National Laboratory, 1 Cyclotron Road, Berkeley, CA 94720, USA}

\collaboration{1000}{(DES Collaboration)}

\email{keagraha@ucsc.edu, tesla@ucsc.edu}

\begin{abstract}
We search for the presence of cool cores in optically-selected galaxy clusters from the Dark Energy Survey (DES) and investigate their prevalence as a function of redshift and cluster richness.  Clusters were selected from the redMaPPer analysis of three years of DES observations that have archival Chandra X-ray observations, giving a sample of 99 clusters with a redshift range of $0.11 < z < 0.87$ and a richness range of $25 < \lambda < 207$. Using the X-ray data, the core temperature was compared to the outer temperature to identify clusters where the core temperature is a factor of 0.7 or less than the outer temperature.  We found a cool core fraction of approximately 20\% with no significant trend in the cool core fraction with either redshift or richness.
\end{abstract}

\section{Introduction}
Galaxy clusters and the intra-cluster medium (ICM) are excellent probes of the evolution of large-scale structure. Understanding the thermodynamics of clusters and the ICM is essential to understanding their growth. The X-ray emitting ICM in the center of clusters is dense enough that it should be able to radiatively cool in less than the Hubble time \citep[e.g.][]{Fabian94,White97}. This central cooling gas would be compressed by the external, hotter gas creating a ``cooling flow". However, observations show that the central gas does not cool all the way down producing the expected star formation and neutral gas, forming instead a ``cool core" and indicating the presence of some form of feedback such as AGN feedback that prevents the cluster from fully cooling \citep[e.g.][]{peterson06}. Cool-core clusters are of interest as understanding their evolution may help understand the processes like cluster mergers and AGN feedback which can disrupt cooling and provide sources of heating \citep[e.g.][]{Voit05,mcnamara07,Hudson10,fabian12,mcnamara12,McDonald18}.  

One might expect trends in the level of cooling and the fraction of clusters with cool cores as a function of redshift and cluster mass \citep[e.g][]{Santos10, mcdonald13, pascut15}. Redshift trends in the cool-core fraction will depend on factors such as the possibly evolving level of AGN feedback, the AGN on-off duty cycle and its relation to cooling, the cluster merger rate, the extent to which mergers disrupt cooling, and the post merger relaxation timescale \citep[e.g.][]{Henning09,Gaspari11}.
The evolution of cool core clusters as a function of mass is also of interest. As the gas in low mass clusters and groups is less gravitationally bound, AGN feedback can be more effective \citep[e.g.][]{McCarthy10,Eckert21}, and in fact strong cool cores are not observed in groups hosting a strong, radio-loud AGN, opposite the trend seen for clusters \citep{Sun09b,Bharadwaj14}.

This paper examines the cool core fraction in a sample of 99 optically-selected clusters as a function of redshift and richness. We also investigate the distributions of the ratio of core temperature to the temperature of the cluster at larger radii ignoring the core in bins of redshift and richness. 

\section{Data Analysis and Results}
Our sample is drawn from Chandra observations of DES clusters selected by the redMaPPer algorithm \citep{Rykoff14}, specifically the Y3 6.4.22+2 catalog. We used the Mass Analysis Tool for Chandra \citep[MATCha,][]{Hollowood19} pipeline to determine cluster X-ray temperatures and luminosities, as well as to perform centering measurements for the redMaPPer clusters which have archival Chandra data \citep{Kelly22}. In this work, we use an updated version of MATCha that fits the core temperature of clusters within $0.15 r_{500}$ of the X-ray peak as well as core-cropped $r_{500}$ temperature, where $r_{500}$ is the radius at which the density is 500 times the critical density. Clusters with observations of insufficient depth such that the core temperature or $r_{500}$ core-cropped temperature fits failed were not included; we also removed clusters that were flagged as contaminated by a foreground/background cluster, had Chandra data taken in a non-imaging mode, or had a signal-to-noise ratio within a radius of 500 kpc less than 9 \citep[see][]{Hollowood19,Kelly22}. With these filters, we were left with a sample of 99 clusters with $0.11 < z < 0.87$ and $25 < \lambda < 207$. This sample size is similar to previous works that analyzed cool core fraction using X-ray data \citep[e.g][]{Santos10, McDonald18, ruppin21}.

To classify cool-core clusters, we measured the ratio of core temperature within $0.15 r_{500}$ to the $r_{500}$ core-cropped temperature ($T_{\mathrm{core}}/T_{r500}$). Core temperatures  
had an average uncertainty of approximately 20\%; temperature errors had a negligible effect on the derived cool-core fractions compared to the statistical errors given the limited sample size.  Any cluster with a ratio of core to non-core temperature less than 0.7 was defined to be a cool core cluster. This definition is a simplification of the definition in \citet{Hudson10} whose well-sampled data allowed fitting of the temperature profile to the virial radius and determination of the core size. 
By this definition, our sample contained 21 cool-core clusters and 78 non cool-core clusters. We looked at the distributions of $T_{\mathrm{core}}/T_{r500}$ in bins of redshift and richness to see if cooling occurs at comparable levels. We used two bins for each with nominal splits at $z = 0.45$ for redshift and $\lambda = 100$ for richness, roughly the means of the sample’s redshift and richness, respectively. These split values were varied, but had little to no impact on the resulting distributions. The nominal redshift bins had 32 high-redshift clusters and 67 low-redshift clusters, and the bins of richness had 54 high-richness clusters and 45 low-richness clusters.

Histograms of $T_{\mathrm{core}}/T_{r500}$ are shown in Figure 1 for the bins of low versus high redshifts (left) and low versus high richnesses (right). As can be seen in the figure, the distributions of temperature ratio are similar for both the redshift and richness bins.  To quantify this we ran a two-sample Kolmogorov-Smirnov Test (KS Test) on both sets of distributions. For the redshift bins, the KS test gave a p-value of 
 0.74 while for the bins of richness the p-value was 0.75, indicating no significant difference in both cases.

\begin{figure}
\includegraphics[width=\linewidth]{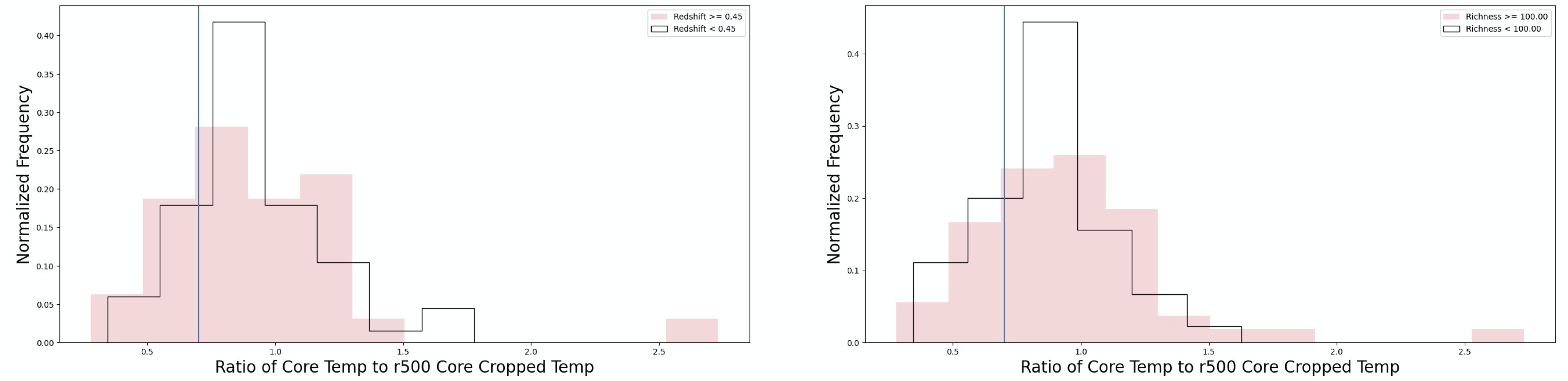}
\caption{\textit{Left:} Histograms of the distributions of $T_{\mathrm{core}}/T_{r500}$ for clusters with $z \ge 0.45$ (pink) and $z<0.45$ (blue). The blue line indicates a temperature ratio of 0.7 below which is our adopted definition of cool core clusters. \textit{Right:} Histograms of the distributions of core to outer temperature for clusters with $\lambda \ge 100$ (pink) and $\lambda< 100$ (blue); the blue vertical line again indicates our adopted cool-core cut.
}
\end{figure}

We also calculated the fraction of cool cores in each redshift and richness bin. This resulted in a cool-core fraction of $0.25 \pm 0.09$ for high redshifts and $0.19 \pm 0.05$ for low redshifts, which are consistent within the uncertainties. Uncertainties were estimated by taking the square root of the number of cool cores in each bin. We also varied the redshift split to see if these fractions changed significantly, but in all cases the cool-core fraction was found to be roughly 0.2 indicating little dependence on redshift. 
The cool-core fraction for high richness was found to be $0.22 \pm 0.06$ while for low richness it was found to be $0.20 \pm 0.08$. The richness split was also varied, but again we obtained similar results with a cool-core fraction around 0.2 in all cases.

\section{Conclusion}
We find a cool-core fraction of $\sim 20$\% of clusters and distributions of $T_{\mathrm{core}}/T_{r500}$ which are statistically consistent in bins of redshift and richness.  While the distribution of cool cores is fairly even throughout our sample, this could be due to limited data. Our sample lacked a significant number of clusters with fairly high redshifts ($z  > 0.7$) and with fairly low richnesses ($\lambda <60$). It should be noted as well that X-ray surveys have been known to be biased towards cool cores when compared to SZ surveys \citep[see][]{Rossetti16,Andrade-Santos17}, although \citet{Ghirardini22} do not find a bias toward cool cores in the eROSITA eFEDS cluster sample. In addition, our clusters are primarily drawn from targeted Chandra observations possibly introducing a bias. Despite this, our results are generally consistent with previous works. Previous results on the redshift evolution in the cool core fraction and the properties of cool cores are mixed with some studies finding a decreasing fraction of cool cores with increasing redshift \citep{Santos10,pascut15,Ghirardini22} and others finding no evolution \citep{McDonald17,sanders18,ruppin21}. The exact redshift trends depend on the cool core metric used and the sample selection \citep[e.g.][]{mcdonald13,pascut15}; the lack of evolution we see in core temperature drop is consistent with observations showing a lack of evolution in cooling properties. However, evolution does occur in cluster surface brightness profiles due to the build up of the bulk cluster around a fixed core \citep[e.g.][]{McDonald17,ruppin21}.
On the other hand, previous work has indicated that cool-core clusters form with little dependence on mass \citep[e.g.][]{Bharadwaj14,pascut15} consistent with the lack of trend with richness in our sample, though important differences are seen between groups and clusters in the relationship of central AGN to the presence of a cool core \citep{Sun09b,McDonald11,Bharadwaj14}. 

\section*{Acknowledgements}
K.G. gratefully acknowledges support from the Koret Scholars Program. This work was supported by the U.S. Department of Energy under Award Number DE-SC0010107. Funding for the DES Projects has been provided by the DOE and NSF(USA), MEC/MICINN/MINECO(Spain), STFC(UK), HEFCE(UK). NCSA(UIUC), KICP(U. Chicago), CCAPP(Ohio State), 
MIFPA(Texas A\&M), CNPQ, FAPERJ, FINEP (Brazil), DFG(Germany) and the Collaborating Institutions in the Dark Energy Survey.

The Collaborating Institutions are Argonne Lab, UC Santa Cruz, University of Cambridge, CIEMAT-Madrid, University of Chicago, University College London, DES-Brazil Consortium, University of Edinburgh, ETH Z{\"u}rich, Fermilab, University of Illinois, ICE (IEEC-CSIC), IFAE Barcelona, Lawrence Berkeley Lab, LMU M{\"u}nchen and the associated Excellence Cluster Universe, University of Michigan, NFS's NOIRLab, University of Nottingham, Ohio State University, University of 
Pennsylvania, University of Portsmouth, SLAC National Lab, Stanford University, University of Sussex, Texas A\&M University, and the OzDES Membership Consortium.

Based in part on observations at Cerro Tololo Inter-American Observatory at NSF's NOIRLab (NOIRLab Prop. ID 2012B-0001; PI: J. Frieman), which is managed by the Association of Universities for Research in Astronomy (AURA) under a cooperative agreement with the National Science Foundation.

The DES Data Management System is supported by the NSF under Grant Numbers AST-1138766 and AST-1536171. 
The DES participants from Spanish institutions are partially supported by MICINN under grants ESP2017-89838, PGC2018-094773, PGC2018-102021, SEV-2016-0588, SEV-2016-0597, and MDM-2015-0509, some of which include ERDF funds from the European Union. IFAE is partially funded by the CERCA program of the Generalitat de Catalunya.
Research leading to these results has received funding from the European Research Council under the European Union's Seventh Framework Program (FP7/2007-2013) including ERC grant agreements 240672, 291329, and 306478.
We  acknowledge support from the Brazilian Instituto Nacional de Ci\^encia
e Tecnologia (INCT) do e-Universo (CNPq grant 465376/2014-2).

This manuscript has been authored by Fermi Research Alliance, LLC under Contract No. DE-AC02-07CH11359 with the U.S. Department of Energy, Office of Science, Office of High Energy Physics. 

\bibliography{refs}{}

\begin{thebibliography}{}
\expandafter\ifx\csname natexlab\endcsname\relax\def\natexlab#1{#1}\fi
\providecommand{\url}[1]{\href{#1}{#1}}
\providecommand{\dodoi}[1]{doi:~\href{http://doi.org/#1}{\nolinkurl{#1}}}
\providecommand{\doeprint}[1]{\href{http://ascl.net/#1}{\nolinkurl{http://ascl.net/#1}}}
\providecommand{\doarXiv}[1]{\href{https://arxiv.org/abs/#1}{\nolinkurl{https://arxiv.org/abs/#1}}}

\bibitem[{Andrade-Santos {et~al.}(2017)Andrade-Santos, Jones, Forman, Lovisari,
  Vikhlinin, van Weeren, Murray, Arnaud, Pratt, Démoclès, Kraft, Mazzotta,
  Böhringer, Chon, Giacintucci, Clarke, Borgani, David, Douspis,
  Pointecouteau, Dahle, Brown, Aghanim, \& Rasia}]{Andrade-Santos17}
Andrade-Santos, F., Jones, C., Forman, W.~R., {et~al.} 2017, The Astrophysical
  Journal, 843, 76, \dodoi{10.3847/1538-4357/aa7461}

\bibitem[{{Bharadwaj} {et~al.}(2014){Bharadwaj}, {Reiprich}, {Schellenberger},
  {Eckmiller}, {Mittal}, \& {Israel}}]{Bharadwaj14}
{Bharadwaj}, V., {Reiprich}, T.~H., {Schellenberger}, G., {et~al.} 2014, \aap,
  572, A46, \dodoi{10.1051/0004-6361/201322684}

\bibitem[{{Eckert} {et~al.}(2021){Eckert}, {Gaspari}, {Gastaldello}, {Le Brun},
  \& {O'Sullivan}}]{Eckert21}
{Eckert}, D., {Gaspari}, M., {Gastaldello}, F., {Le Brun}, A. M.~C., \&
  {O'Sullivan}, E. 2021, Universe, 7, 142, \dodoi{10.3390/universe7050142}

\bibitem[{{Fabian}(1994)}]{Fabian94}
{Fabian}, A.~C. 1994, \araa, 32, 277,
  \dodoi{10.1146/annurev.aa.32.090194.001425}

\bibitem[{{Fabian}(2012)}]{fabian12}
---. 2012, \araa, 50, 455, \dodoi{10.1146/annurev-astro-081811-125521}

\bibitem[{{Gaspari} {et~al.}(2011){Gaspari}, {Melioli}, {Brighenti}, \&
  {D'Ercole}}]{Gaspari11}
{Gaspari}, M., {Melioli}, C., {Brighenti}, F., \& {D'Ercole}, A. 2011, \mnras,
  411, 349, \dodoi{10.1111/j.1365-2966.2010.17688.x}

\bibitem[{Ghirardini {et~al.}(2022)Ghirardini, Bahar, Bulbul, Liu, Clerc,
  Pacaud, Comparat, Liu, Ramos-Ceja, Hoang, Ider-Chitham, Klein, Merloni,
  Nandra, Ota, Predehl, Reiprich, Sanders, \& Schrabback}]{Ghirardini22}
Ghirardini, V., Bahar, Y.~E., Bulbul, E., {et~al.} 2022, Astronomy {\&};
  Astrophysics, 661, A12, \dodoi{10.1051/0004-6361/202141639}

\bibitem[{{Henning} {et~al.}(2009){Henning}, {Gantner}, {Burns}, \&
  {Hallman}}]{Henning09}
{Henning}, J.~W., {Gantner}, B., {Burns}, J.~O., \& {Hallman}, E.~J. 2009,
  \apj, 697, 1597, \dodoi{10.1088/0004-637X/697/2/1597}

\bibitem[{{Hollowood} {et~al.}(2019){Hollowood}, {Jeltema}, {Chen}, {Farahi},
  {Evrard}, {Everett}, {Rozo}, {Rykoff}, {Bernstein}, {Bermeo-Hernandez},
  {Eiger}, {Giles}, {Israel}, {Michel}, {Noorali}, {Romer}, {Rooney}, \&
  {Splettstoesser}}]{Hollowood19}
{Hollowood}, D.~L., {Jeltema}, T., {Chen}, X., {et~al.} 2019, \apjs, 244, 22,
  \dodoi{10.3847/1538-4365/ab3d27}

\bibitem[{{Hudson} {et~al.}(2010){Hudson}, {Mittal}, {Reiprich}, {Nulsen},
  {Andernach}, \& {Sarazin}}]{Hudson10}
{Hudson}, D.~S., {Mittal}, R., {Reiprich}, T.~H., {et~al.} 2010, \aap, 513,
  A37, \dodoi{10.1051/0004-6361/200912377}

\bibitem[{{Kelly} {et~al.}(in prep.){Kelly}, {Jobel}, {Eiger}, {Hollowood},
  {Abd}, {Jeltema}, {Giles}, {Bhargava}, {Everett}, \& {Wilkinson}}]{Kelly22}
{Kelly}, P., {Jobel}, J., {Eiger}, O., {et~al.} in prep.

\bibitem[{{McCarthy} {et~al.}(2010){McCarthy}, {Schaye}, {Ponman}, {Bower},
  {Booth}, {Dalla Vecchia}, {Crain}, {Springel}, {Theuns}, \&
  {Wiersma}}]{McCarthy10}
{McCarthy}, I.~G., {Schaye}, J., {Ponman}, T.~J., {et~al.} 2010, \mnras, 406,
  822, \dodoi{10.1111/j.1365-2966.2010.16750.x}

\bibitem[{{McDonald} {et~al.}(2018){McDonald}, {Gaspari}, {McNamara}, \&
  {Tremblay}}]{McDonald18}
{McDonald}, M., {Gaspari}, M., {McNamara}, B.~R., \& {Tremblay}, G.~R. 2018,
  \apj, 858, 45, \dodoi{10.3847/1538-4357/aabace}

\bibitem[{{McDonald} {et~al.}(2011){McDonald}, {Veilleux}, \&
  {Mushotzky}}]{McDonald11}
{McDonald}, M., {Veilleux}, S., \& {Mushotzky}, R. 2011, \apj, 731, 33,
  \dodoi{10.1088/0004-637X/731/1/33}

\bibitem[{{McDonald} {et~al.}(2013){McDonald}, {Benson}, {Vikhlinin},
  {Stalder}, {Bleem}, {de Haan}, {Lin}, {Aird}, {Ashby}, {Bautz}, {Bayliss},
  {Bocquet}, {Brodwin}, {Carlstrom}, {Chang}, {Cho}, {Clocchiatti}, {Crawford},
  {Crites}, {Desai}, {Dobbs}, {Dudley}, {Foley}, {Forman}, {George},
  {Gettings}, {Gladders}, {Gonzalez}, {Halverson}, {High}, {Holder},
  {Holzapfel}, {Hoover}, {Hrubes}, {Jones}, {Joy}, {Keisler}, {Knox}, {Lee},
  {Leitch}, {Liu}, {Lueker}, {Luong-Van}, {Mantz}, {Marrone}, {McMahon},
  {Mehl}, {Meyer}, {Miller}, {Mocanu}, {Mohr}, {Montroy}, {Murray},
  {Nurgaliev}, {Padin}, {Plagge}, {Pryke}, {Reichardt}, {Rest}, {Ruel}, {Ruhl},
  {Saliwanchik}, {Saro}, {Sayre}, {Schaffer}, {Shirokoff}, {Song},
  {{\v{S}}uhada}, {Spieler}, {Stanford}, {Staniszewski}, {Stark}, {Story}, {van
  Engelen}, {Vanderlinde}, {Vieira}, {Williamson}, {Zahn}, \&
  {Zenteno}}]{mcdonald13}
{McDonald}, M., {Benson}, B.~A., {Vikhlinin}, A., {et~al.} 2013, \apj, 774, 23,
  \dodoi{10.1088/0004-637X/774/1/23}

\bibitem[{{McDonald} {et~al.}(2017){McDonald}, {Allen}, {Bayliss}, {Benson},
  {Bleem}, {Brodwin}, {Bulbul}, {Carlstrom}, {Forman}, {Hlavacek-Larrondo},
  {Garmire}, {Gaspari}, {Gladders}, {Mantz}, \& {Murray}}]{McDonald17}
{McDonald}, M., {Allen}, S.~W., {Bayliss}, M., {et~al.} 2017, \apj, 843, 28,
  \dodoi{10.3847/1538-4357/aa7740}

\bibitem[{{McNamara} \& {Nulsen}(2007)}]{mcnamara07}
{McNamara}, B.~R., \& {Nulsen}, P.~E.~J. 2007, \araa, 45, 117,
  \dodoi{10.1146/annurev.astro.45.051806.110625}

\bibitem[{{McNamara} \& {Nulsen}(2012)}]{mcnamara12}
---. 2012, New Journal of Physics, 14, 055023,
  \dodoi{10.1088/1367-2630/14/5/055023}

\bibitem[{{Pascut} \& {Ponman}(2015)}]{pascut15}
{Pascut}, A., \& {Ponman}, T.~J. 2015, \mnras, 447, 3723,
  \dodoi{10.1093/mnras/stu2688}

\bibitem[{{Peterson} \& {Fabian}(2006)}]{peterson06}
{Peterson}, J.~R., \& {Fabian}, A.~C. 2006, \physrep, 427, 1,
  \dodoi{10.1016/j.physrep.2005.12.007}

\bibitem[{{Rossetti} {et~al.}(2016){Rossetti}, {Gastaldello}, {Ferioli},
  {Bersanelli}, {De Grandi}, {Eckert}, {Ghizzardi}, {Maino}, \&
  {Molendi}}]{Rossetti16}
{Rossetti}, M., {Gastaldello}, F., {Ferioli}, G., {et~al.} 2016, \mnras, 457,
  4515, \dodoi{10.1093/mnras/stw265}

\bibitem[{{Ruppin} {et~al.}(2021){Ruppin}, {McDonald}, {Bleem}, {Allen},
  {Benson}, {Calzadilla}, {Khullar}, \& {Floyd}}]{ruppin21}
{Ruppin}, F., {McDonald}, M., {Bleem}, L.~E., {et~al.} 2021, \apj, 918, 43,
  \dodoi{10.3847/1538-4357/ac0bba}

\bibitem[{{Rykoff} {et~al.}(2014){Rykoff}, {Rozo}, {Busha}, {Cunha},
  {Finoguenov}, {Evrard}, {Hao}, {Koester}, {Leauthaud}, {Nord}, {Pierre},
  {Reddick}, {Sadibekova}, {Sheldon}, \& {Wechsler}}]{Rykoff14}
{Rykoff}, E.~S., {Rozo}, E., {Busha}, M.~T., {et~al.} 2014, \apj, 785, 104,
  \dodoi{10.1088/0004-637X/785/2/104}

\bibitem[{{Sanders} {et~al.}(2018){Sanders}, {Fabian}, {Russell}, \&
  {Walker}}]{sanders18}
{Sanders}, J.~S., {Fabian}, A.~C., {Russell}, H.~R., \& {Walker}, S.~A. 2018,
  \mnras, 474, 1065, \dodoi{10.1093/mnras/stx2796}

\bibitem[{{Santos} {et~al.}(2010){Santos}, {Tozzi}, {Rosati}, \&
  {B{\"o}hringer}}]{Santos10}
{Santos}, J.~S., {Tozzi}, P., {Rosati}, P., \& {B{\"o}hringer}, H. 2010, \aap,
  521, A64, \dodoi{10.1051/0004-6361/201015208}

\bibitem[{{Sun}(2009)}]{Sun09b}
{Sun}, M. 2009, \apj, 704, 1586, \dodoi{10.1088/0004-637X/704/2/1586}

\bibitem[{{Voit}(2005)}]{Voit05}
{Voit}, G.~M. 2005, Reviews of Modern Physics, 77, 207,
  \dodoi{10.1103/RevModPhys.77.207}

\bibitem[{{White} {et~al.}(1997){White}, {Jones}, \& {Forman}}]{White97}
{White}, D.~A., {Jones}, C., \& {Forman}, W. 1997, \mnras, 292, 419,
  \dodoi{10.1093/mnras/292.2.419}

\end{thebibliography}
\bibliographystyle{aasjournal}

\end{document}